# Quasiparticle energies and excitonic effects of α-RuCl$_3$

*Du Li[1], and Li Yang[1,2, *]*

[1]Department of Physics, Washington University in St. Louis, St. Louis, Missouri 63130, USA

[2]Institute of Materials Science and Engineering, Washington University in St. Louis, St. Louis, Missouri 63130, USA

*Correspondence author: lyang@physics.wustl.edu

**ABSTRACT**

α-phase Ruthenium (III) chloride (α-RuCl$_3$) has attracted significant attention because of its potential for realizing Kitaev quantum spin liquid. In this work, we employ first-principles many-body perturbation theory to study its many-electron interactions and excited-state properties. We find enhanced many-electron interactions that dominate quasiparticle energies and optical responses in α-RuCl$_3$. Our calculated quasiparticle band gap of bulk structure is about 1.75 eV that agree well with recent scanning tunneling spectroscopy and angle-resolved photoemission spectroscopy measurements. Furthermore, our calculated first and second primary bright excitons are located at 1.23 eV and 1.98 eV, respectively. These excitons show good consistency with the main features observed in the optical absorption spectrum. We extend our investigation to monolayer α-RuCl$_3$, examining the zigzag antiferromagnetic (AFM) and ferromagnetic (FM) phases. In addition to significant excitonic effect., the optical spectrum of the zigzag AFM phase exhibits anisotropic behavior, while the FM phase demonstrates isotropic characteristics. The different optical response behaviors provide an efficient approach to identify the energy nearly degenerate magnetic states, which can both potentially exist in fabricated samples.

# I. INTRODUCTION

Among emerging van der Waals (vdW) magnetic materials, α-phase Ruthenium (III) chloride (α-RuCl$_3$) has attracted significant research interest. Bulk α-RuCl$_3$ has a "zigzag" antiferromagnetic (zigzag AFM) order with a Neel temperature of 7 K [1-4] and is an excellent platform for investigating opto-magnetic effects [5]. In the temperature range of 7 to 120 K, experimental evidence for proximate Kitaev quantum spin liquid states and Majorana fermions has been observed through various measurements, including neutron scattering [4,6] and Raman spectroscopy [7-10], etc. Substantial efforts have recently been made to achieve ultra-thin α-RuCl$_3$ [11-13], which is anticipated to serve as an ideal platform for the realization of the two-dimensional (2D) Kitaev honeycomb model [14-16].

However, many fundamental excited-state properties of α-RuCl$_3$ remain ambiguous. The physics nature of the electronic gap and the origin of the primary optical features are still elusive. Intensive angle-resolved photoemission spectroscopy (ARPES) and scanning tunneling spectroscopy (STS) measurements have been performed but resulted in diverse conclusions about the quasiparticle (QP) band gaps: 0.2-0.3 eV, above 1.2 eV, and around 1.8 eV [17-20]. In contrast, the optical absorption studies also have identified two energy ranges for the optical gap: approximately 0.2 eV and around 1.0 eV [21-24]. Meanwhile, the density functional theory (DFT) plus Hubbard U calculation is a widely used method to study the electronic structure. To determine the essential Hubbard parameters, Kee's group analyzed the x-ray absorption spectroscopy and optical features in the absorption spectrum using the single-ion model and got the Hubbard parameters U = 2.4 eV and J = 0.4 eV [23,24]. The DFT+U calculated band dispersions agreed well with the ARPES measurements near the band edge [18,19,24], but the bandgap value (~1.0 eV) is significantly smaller than the ARPES and STS measurements (~1.8 eV). This indicates more sophisticated physics nature need to be considered to fully understand the electronic structure and excited-state properties.

Notably, many-electron interactions are known important in deciding excited-state properties of insulators. Optically excited electron-hole pairs and excitonic effects may introduce significant corrections to optical responses based on the single-particle picture. More specifically, both DFT+U calculations and ARPES measurements [18,19,24] reveal the presence of flat bands near the band edge in α-RuCl$_3$. These flat bands contribute significantly to a large joint density of states (JDOS), which may further enhance the chance of forming electron-hole pairs. Consequently, all these factors collectively suggest that many-electron interactions are crucial for a comprehensive understanding of both the QP electronic structure and the excitonic optical characteristics of α-RuCl$_3$, which remains elusive.

In this work, we employed first-principles many-body perturbation theory (MBPT) to systematically study the many-electron interactions and excited-state properties of bulk and 2D α-RuCl$_3$. We calculated the QP energies within the *GW* approximation and investigated the excitonic effects by solving the Bethe-Salpeter equation (BSE). For bulk α-RuCl$_3$, our calculated QP band gap of 1.75 eV agrees well with recent STS and ARPES measurements [19,20]. Our calculated optical absorption spectrum, which includes excitonic effects, shows good consistency with the major profile of experimental results [23,24]. Furthermore, we have calculated the QP energies and the absorption spectrum for monolayer 2D α-RuCl$_3$. The self-energy correction and excitonic effects are substantially enhanced due to 2D quantum confinement and reduced screening. The calculated QP band gap (~ 2.20 eV) agrees satisfactorily with recent STS measurements (above 2.00 eV) [13]. Moreover, our BSE calculations revealed significant optical signatures to differentiate the energy nearly degenerate magnetic orders, specifically the zigzag AFM and ferromagnetic (FM) phases. The optical spectrum of the zigzag AFM phase exhibits anisotropic behavior, while that of the FM phase is isotropic. These results provide an efficient optical approach to identify ultra-thin frustrated magnets and to explore the potential quantum spin liquid state. Our studies reveal the underlying physics arising from the interplay between quasiparticle energies and excitonic effects in α-RuCl$_3$. The observed agreements and discrepancies in excitonic behavior and optical absorption measurements provide fundamental insights and motivation for further investigations into correlated materials, a class of systems of central importance in condensed matter physics.

The remainder of this article is organized as follows: In Sec. II, we present the computational details of the simulation set up. The results of the electronic and optical properties for bulk and monolayer α-RuCl$_3$ are presented In Secs. III, IV, respectively. Finally, the conclusions are summarized in Sec. V.

## II. COMPUTATIONAL DETAILS

The DFT calculation is performed using the PBE exchange−correlation functional [25,26], as implemented in the Quantum ESPRESSO package [27]. A 1 × 2 × 1 supercell is used to construct the in-plane zigzag AFM magnetic orders. A wavefunction cutoff of 60 Ry (detailed in the convergence test in Supplementary Material Sec. I [28]) and a charge density cutoff of 240 Ry are adopted. Hubbard parameters U = 2.4 eV and J = 0.4 eV [23,24,30] are used for the Ruthenium atom. The van der Waals (vdW) interaction is included via the semiempirical Grimme-D3 scheme [31]. A vacuum of 18 Å between adjacent layers is used to avoid spurious interactions between periodic images along the out-of-plane direction of 2D monolayer structures. Our calculation shows that spin orbital coupling (SOC) does not significantly influence the primary optical features interested in this work. Thus, it is not included in the simulations (as discussed in Supplementary Materials Sec. II [28]).

The *GW* calculations are performed using the BerkeleyGW package [32]. All *GW* calculations reported in this work are perturbative, i.e., the single-shot $G_0W_0$. The Hybertsen−Louie generalized plasmon-pole model is used to treat the frequency dependence of the dielectric function [33]. The static remainder approximation [34] is used in the evaluation of the self-energy for faster convergence. The treatment of Hubbard U term in Dyson equation is detailed in Supplementary Material Sec. V [28]. The Bethe-Salpeter equation (BSE) is employed to obtain excitonic effects and optical absorption spectra [35]. The slab Coulomb truncation [36] is adopted to remove the spurious long-range Coulomb interactions along the non-periodic direction for 2D structures. The coarse k grid is set to be 9 × 4 × 4 for calculating the dielectric function and QP energies, and it is then interpolated to a fine k grid of 18 × 8 × 8 for computing the electron-hole interaction kernel and solving the BSE. For monolayer zigzag AFM $RuCl_3$, a coarse k grid of 9 × 4 × 1 is used, and an 18 × 8 × 1 k grid is adopted as the fine k grid. In the case of monolayer FM $RuCl_3$, the coarse and fine k grids are set to be 9 × 9 × 1 and 18 × 18 × 1, respectively. The convergence tests are shown in Supplementary Material Sec. I [28]. Gaussian curves with an energy resolution of 0.08 eV are adopted to broaden the exciton peaks in the calculated absorption spectrum. And a linear fit of the QP energy corrections (as shown in Supplementary Material Sec. III [28]) is used for the fine grids. The convergences of QP energies and excitons are within 80 meV.

## III. Bulk α-RuCl₃

We start by considering the crystal structure and magnetic properties of bulk α-RuCl$_3$. Fig. 1(a) shows the in-plane crystal structure, where α-RuCl$_3$ forms a monoclinic C2/m space group, featuring a symmetrical honeycomb lattice. Ru$^{3+}$ ions are coordinated with six Cl$^-$ ions, forming an edge-sharing RuCl$_6$ octahedra. Our calculated in-plane lattice constant for α-RuCl$_3$ is 5.99 Å, which agree well with the experimental value within 1% [2,3]. This honeycomb lattice consists of two interpenetrating triangular sublattices. The terms zigzag direction and armchair direction describe distinct orientations of atomic chains within this structure. The zigzag direction (defined as the x-direction) consists of atomic chains connected at alternating angles, forming a characteristic zigzag pattern. In contrast, the armchair direction (defined as the y-direction) is perpendicular to the zigzag direction and consists of atomic rows that resemble an armchair-like shape when viewed along this axis.

The intralayer magnetic structure exhibits a zigzag AFM order [1-4]. In this configuration, Ru$^{3+}$ ions align their magnetic moments ferromagnetically within the same zigzag chain, while maintaining an AFM alignment with neighboring chains. Perpendicular to the plane, α-RuCl$_3$ displays a layered structure with tilted octahedra. Each subsequent layer is shifted by a/3 relative to the layer below as depicted in Fig. 1(b), resulting in a periodic ABC stacking where every fourth layer aligns directly above the first layer. Although the interlayer magnetic order is crucial for understanding the magnetic properties, our simulations indicates that it does not significantly affect the electronic band gap and primary optical properties, as detailed in the Supplementary Material Sec. IV [28]. Consequently, we focus on the interlayer FM magnetic order in the following discussions for bulk α-RuCl$_3$.

The DFT-calculated band structure of bulk α-RuCl$_3$ is presented in Fig. 1(c). The valence band maximum (VBM) is located at the high-symmetry M point and the conduction band minimum (CBM) is located at the Γ point. This reveals an indirect M-Γ band gap of 0.77 eV and a direct Γ-Γ band gap of 0.78 eV. We have calculated QP electronic structures within the *GW* approximation as represented in Fig 1(d). The DFT and QP band gaps are summarized in Table I. The indirect M-Γ QP band gap is 1.75 eV and a direct Γ-Γ QP band gap is 1.77 eV. The QP energy corrections in bulk α-RuCl$_3$ are substantial, rendering a 0.98 eV enlargement of the DFT result. This enlargement of QP band gap is more pronounced than those in bulk layered transition metal dichalcogenide (TMD) materials (∼0.3-0.4 eV) [37] and black phosphorus (less than 0.3 eV) [38]. The reason for this difference lies in the relatively flat dispersions exhibited by both the valence and conduction band edges, as illustrated in Fig. 1(d). This increased DOS is indicative of stronger many-electron interactions, resulting in more significant QP corrections. The similar conclusions are observed in other calculations of 2D magnets [39-41]. Importantly, our calculated QP band gap agrees well with experimental results obtained by the STS and ARPES measurements (around 1.8 eV) [19,20].

The optical absorption spectrum has been calculated both with and without excitonic effects, as illustrated in Fig. 2. Taking the zigzag polarization direction as an example (Fig. 2(a)), when electron-hole interactions are not considered, the optical absorption (represented by the blue dashed line) begins at approximately 1.8 eV. It is noteworthy that this value is slightly larger than the QP band gap. This is because the dipole oscillator strength arising from the band edge transitions is significantly weaker, about 2-3 orders of magnitude smaller, compared to the transitions involving the first two conduction bands and the inner valence bands. The absorption spectrum exhibits two prominent peaks. The first major peak, observed at 2.0 eV, is primarily attributed to transitions between the first and second conduction bands and the third and fourth valence bands. The second significant peak, occurring around 2.4 eV, mainly results from transitions between the first and second conduction bands and the fifth and sixth valence bands.

The incorporation of electron-hole interactions significantly alters the optical absorption spectrum. Notably, α-RuCl$_3$ exhibits anisotropic optical characteristics. For the incident light polarized along the zigzag direction (Fig. 2(a)), the first prominent $α$ peak is positioned at 1.23 eV. Along the armchair polarization direction (Fig. 2(b)), the first main peak shifts to 1.42 eV, with the exciton dipole oscillator strength decreasing by approximately one order of magnitude. An additional significant peak, labeled $β$, appears at 1.98 eV. Table I summarizes the main excitonic features observed in the spectrum.

In practical experimental settings, the incident light may comprise various polarization directions, and the crystal structure might contain different domains [23,24,42]. To account for this, we considered optical absorption with averaged polarization directions relative to the incident ambient light, as illustrated by the solid red lines in Fig. 2(c). This averaged spectrum reveals two primary peaks: α (1.23 eV) and β (1.98 eV). These α and β peaks align remarkably well with the experimental observations, as indicated by the dotted lines in Fig. 2(c) [23,24]. However, our calculations did not reveal the small γ peak around 0.3 eV observed in the experiments, as shown in Fig. 2(c). This low-energy peak may originate from other mechanisms beyond the *GW*-BSE scope of the intrinsic crystal discussed in this work, e.g., defects, multiplet orbital excitations [24] or surface modulations [13].

To obtain a deeper understanding of the excitonic states in α-RuCl$_3$, we investigated the exciton dipole oscillator strength, as illustrated in Fig. 3(a). The lowest-energy exciton is a dark one, labeled as X$_0$ and located at 1.14 eV. This X$_0$ exciton primarily originates from transitions between the first two conduction bands and the first two valence bands, with a binding energy of 0.63 eV. Notably, the exciton dipole oscillator strength of X$_0$ is approximately two orders of magnitude smaller than that of the bright exciton α. This significant difference can be attributed to the relatively dark nature of the involved band-edge transitions. To further elucidate the characteristics of exciton state X$_0$, we investigated the exciton wavefunction $A_{vck}^{X_0}$ solved from the BSE [35] in reciprocal space. In this representation, *v* refers to valence states and *c* refers to conduction states. For illustrative purposes, we used the first conduction band and the first valence band as an example, and it is worth noting that other wavefunctions exhibit similar behaviors. As shown in Fig. 3(b), the X$_0$ exciton wavefunction in reciprocal space is anisotropic and predominantly distributed along the M-Γ high-symmetry line.

Then we turn to the bright α and β excitons. To elucidate their original contributions, we express the exciton optical transition matrix $\langle 0|\hat{v}|\Omega\rangle$ element from the ground (vacuum) state $|0\rangle$ to an exciton state $|\Omega\rangle = \sum_{vck} A_{vck}^{\Omega}|vc\rangle$. Following the approach in Ref. [41,43], we analyze the contribution of single particle interband transition matrix elements $\langle vk|\hat{v}|ck\rangle$ at a specific energy $\omega$:

$$\langle 0|\hat{v}|\Omega\rangle = \sum_{vck} A_{vck}^{\Omega}\langle vk|\hat{v}|ck\rangle = \int S_\Omega(\omega)d\omega, \tag{1}$$

where

$$S_\Omega(\omega) = \sum_{vck} A_{vck}^{\Omega}\langle vk|\hat{v}|ck\rangle \delta(\omega - (E_{ck} - E_{vk})), \tag{2}$$

and

$$I_\Omega(\omega) = \int_0^\omega S_\Omega(\omega')d\omega', \qquad (3)$$

the square of $I_\Omega(\omega)$ is proportional to the exciton dipole oscillator strength. The corresponding interference effect between the single-particle interband transitions and exciton wavefunctions plays a crucial role in determining the overall dipole oscillator strength of excitons. The derivative $dI_\Omega/d\omega$ can reveal the original interband transitions contributions of an exciton. We present $dI_\alpha/d\omega$ for the $\alpha$ exciton in Fig. 3(d) and compare it with the RPA absorption spectrum. A common main peak is observed around 2.0 eV, indicating that the $\alpha$ exciton shares the same original interband transition contributions as the RPA peak at 2.0 eV. These contributions primarily arise from transitions between the first and second conduction bands and the third and fourth valence bands (further analyze is detailed in Supplementary Material Sec. VI [28]). Consequently, the origin of $\alpha$ exciton approximately suggests a binding energy of 0.77 eV, which is larger than 0.63 eV of the $X_0$ exciton. This difference can be explained by the distribution of the exciton wavefunction. Fig. 3(c) illustrates the $\alpha$ exciton wavefunction ($A^\alpha_{vck}$) in reciprocal space, using the first conduction band and the third valence band as an example (other $A^\alpha_{vck}$ components exhibit the similar behavior). In contrast to the $X_0$ exciton wavefunction in Fig. 3(b), the $\alpha$ exciton demonstrates a diffuse distribution of electron-hole pair contributions across the entire Brillouin zone in reciprocal space. This characteristic implies that the $\alpha$ exciton wavefunctions will exhibit a more localized behavior in real space, resulting in a larger binding energy.

Moreover, we also investigate the original interband transitions contributions of the $\beta$ exciton and presents $dI_\beta/d\omega$ in Fig. 3(e). $dI_\beta/d\omega$ exhibits two peaks around 2.0 eV and 2.4 eV, respectively. These two peaks align with the main peaks observed in the RPA spectrum, indicating that the interband transitions associated with both peaks contribute significantly to the formation of the $\beta$ exciton. Consequently, the primary contributions to the $\beta$ exciton arise from interband transitions between the first and second conduction bands and the third, fourth, fifth, and sixth valence bands.

We have discussed the origins of the primary excitons. In addition, magnetic circular dichroism (MCD) is also a powerful tool for exploring the interplay between magnetism, electronic structure, and optical transitions. However, calculating the MCD spectrum demands fully incorporating SOC and external magnetic field in α-RuCl₃ with zigzag AFM configuration. Therefore, detailed MCD analysis remains beyond the scope of the current work but constitutes an important avenue for future studies.

### IV. Monolayer α-RuCl₃

In monolayer α-RuCl₃ case, we have calculated the total energy of different magnetic orders and found that the zigzag AFM phase represents the ground state spin configurations. However, it is noteworthy that the energy difference between the zigzag AFM and FM phases is less than 0.1 meV/Ru (the magnetic orientation is detailed in Supplementary Material Sec. IV [28]). This result is consistent with previous calculations and may be sensitive to the U/J parameters or functionals employed [5]. Given that both phases are possibly observed in fabricated samples [5,44-46], we will consider both zigzag AFM and FM orders in the following calculations for monolayer α-RuCl₃.

The DFT and *GW* calculated QP band gaps of monolayer α-RuCl₃ are summarized in Table I. Fig. 4(a) presents the QP band structures of the zigzag AFM order. The VBM is located at the high-symmetry F point and the CBM is located at the Γ point. This configuration results in an indirect F-Γ QP band gap of 2.20 eV and a direct Γ-Γ QP band gap of 2.21 eV. When compared to the DFT-calculated band gaps (indirect gap of 0.88 eV and direct gap of 0.90 eV), the QP energy corrections in monolayer α-RuCl₃ lead to an enlargement of approximately 1.3 eV in the QP band gap. This correction is more substantial than that observed in the bulk case (around 1 eV), which can be attributed to the reduced dielectric screening effects inherent to the suspended 2D structure. Fig. 4(b) presents the QP band structures of the FM order. The VBM is located at the high-symmetry M point and the CBM is located at the Γ point. The indirect M-Γ QP band gap is 1.84 eV and the direct Γ-Γ QP band gap is 2.06 eV. The STS experiments have indicated that the monolayer α-RuCl₃ exhibit a band gap of approximately 2.0 eV [13]. Notably, there is an energy difference around 0.2 eV between our calculated QP band gap (2.2 eV) and STS experimental result (2.0 eV). This difference may be attributed to environmental screening effects, as the experiments usually utilize a substrate, whereas our current study focuses on freestanding monolayers. The presence of an external substrate can enhance screening and consequently reduce the QP band gap [47-52].

Furthermore, we have calculated the absorption spectrum of monolayer α-RuCl$_3$, as illustrated in Fig. 5. To avoid the artificial effect from the choice of vacuum space in simulations, we plot the optical absorbance by $A(\omega) = \frac{\omega d}{c}\varepsilon_2(\omega)$, where $\varepsilon_2(\omega)$ is the calculated imaginary part of the dielectric function and $d$ represents the distance between adjacent α-RuCl$_3$ layers along the periodic direction of our calculation. Analogous to the bulk case, monolayer α-RuCl$_3$ in the zigzag AFM configuration exhibits anisotropic behavior in its optical spectra, as demonstrated in Figs. 5(a) and (b). Fig. 5(c) presents the averaged absorption spectrum with respect to all polarized directions of the ambient incident light. As expected, the incorporation of excitonic effects significantly alters the absorption spectra. When light is polarized along the zigzag direction (Fig. 5(a)), we observe a prominent α peak at 0.86 eV. In contrast, light polarization along the armchair direction (Fig. 5(b)) reveals a primary β exciton at 1.41 eV. Moreover, we also calculated the absorption spectrum in FM order in Fig. 5(d). In contrast to the anisotropic behavior observed in the zigzag AFM case, the spectrum exhibits isotropic characteristics in the FM configuration. The first and second main bright exciton α and β are located at 0.77 eV and 1.04 eV, respectively. Both magnetic phases can potentially exist in fabricated samples [5,44-46], making their differentiation challenging. The distinct optical responses exhibited by zigzag AFM and FM configurations provide a reliable approach to distinguish the magnetic order.

Notably, for the zigzag AFM configuration, the first dark $X_0$ exciton locates at 0.76 eV with an electron-hole binding energy of 1.45 eV. In comparison, the FM configuration shows that the first dark $X_0$ exciton locates at 0.69 eV with an electron-hole binding energy of 1.37 eV. The binding energy of the $X_0$ exciton in the FM case is approximately 80 meV smaller than that in the zigzag AFM order. This can be attributed to the distinct band structure characteristics. The zigzag AFM conduction band edge displays a flatter dispersion behavior compared to the FM case as shown in Fig. 4. This flatter band structure leads to a larger JDOS and more localized electron-hole pair, resulting in a larger exciton binding energy for the zigzag AFM configuration.

In monolayer α-RuCl$_3$, in addition to optical anisotropy or isotropy, MCD also provides a useful method to distinguish between FM and zigzag AFM phases. The FM phase has a net magnetization and can exhibit MCD without an external magnetic field. To properly study the MCD in the ground state zigzag AFM configuration, both spin–orbit coupling and an external magnetic field need be incorporated, which lies beyond the scope of the present work but constitutes an important avenue for future studies.

## V. CONCLUSION

In this work, we have employed first-principles MBPT within the *GW* approximation to calculate the QP energies. For bulk α-RuCl$_3$, our calculated QP band gap 1.75 eV agrees well with the recent STS and ARPES measurements [19,20]. We further calculate optical absorption spectrum with excitonic effects by solving BSE. The results show a good consistency with the experimental observations [23,24]. We analyze the physics origin of the primary exciton peaks according to the QP interband transitions. These analyses of the primary α and *β* excitons elucidate the interplay between quasiparticle energies and excitonic effects in α-RuCl$_3$. Additionally, the low-energy γ peak, which is observed experimentally but absent in our calculations, suggests physics beyond the standard exciton picture. These agreements and discrepancies provide critical insights and serve as strong motivation for further exploration of correlated materials. Furthermore, we extend our investigation to monolayer α-RuCl$_3$ and calculate the QP band structure and the absorption spectrum. We present a comparative analysis of zigzag AFM and FM phases. Notably, the optical spectrum of the zigzag AFM phase exhibits anisotropic behavior, while the FM phase demonstrates isotropic characteristics. The different optical response behaviors provide an efficient approach to identify the energy nearly degenerate zigzag AFM and FM states, which can both potentially exist in fabricated samples [5,44-46].

**Acknowledgment**

D.L. was supported by the Air Force Office of Scientific Research (AFOSR) grant no. FA9550-22-1-0340, and L.Y. is supported by the National Science Foundation (NSF) grant No. DMR-2124934.

FIGURES:

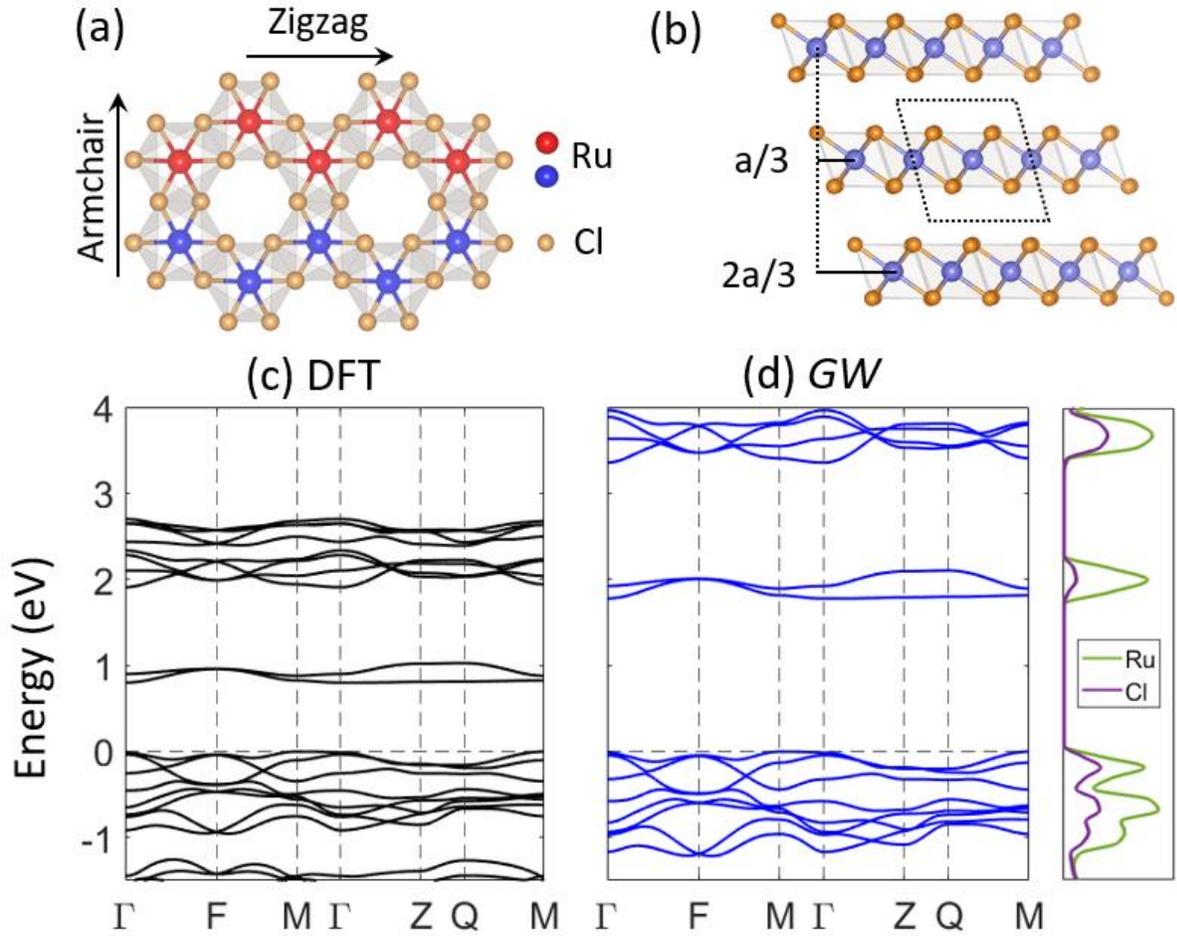

**FIG. 1.** (a) α-RuCl₃ in-plane crystal structure with zigzag AFM magnetic order. Different colors of Ru atoms indicate opposite magnetic orientations. (b) α-RuCl₃ out-of-plane crystal structure with ABC stacking. (c) DFT-calculated electronic structure. (d) The left panel presents QP electronic structure. The right panel presents projected density of states.

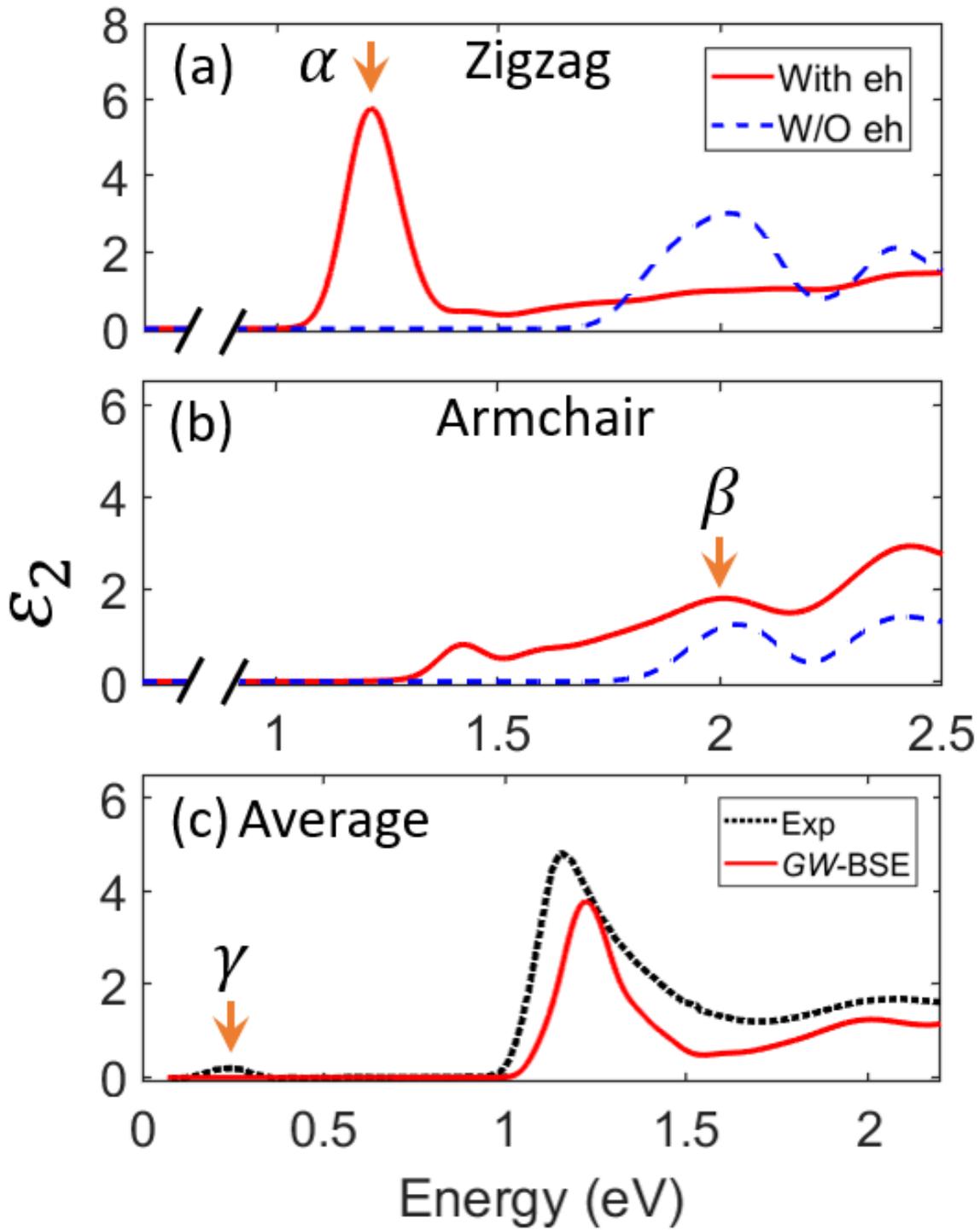

**FIG. 2.** Optical absorption spectrum of bulk α-RuCl$_3$ without electron-hole interaction (blue dashed line) and with electron-hole interaction (red solid line). The incident light is polarized in (a) zigzag direction (b) armchair direction. (c) Absorption spectrum averaging all polarization directions. The dotted black line is experimental data from [24].

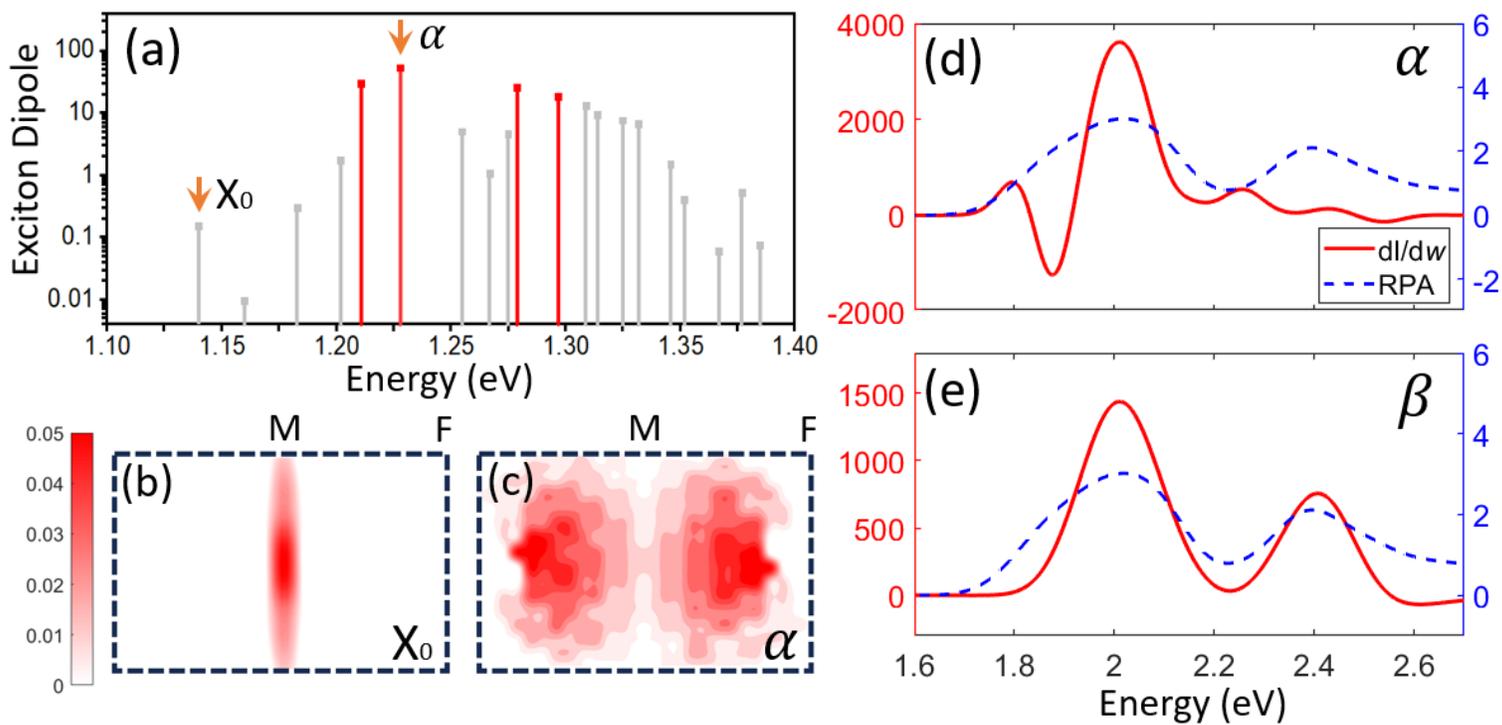

**FIG. 3.** (a) Exciton energy levels versus exciton dipole oscillator strength. Optically bright exciton states are in red while dark exciton states are in grey. Exciton wavefunctions in reciprocal space for (b) $X_0$ and (c) $\alpha$ exciton states (in arbitrary unit). $dI/d\omega$ spectrum (solid red line in arbitrary unit) and $GW$-RPA absorption spectrum for (d) $\alpha$ and (e) $\beta$ exciton states.

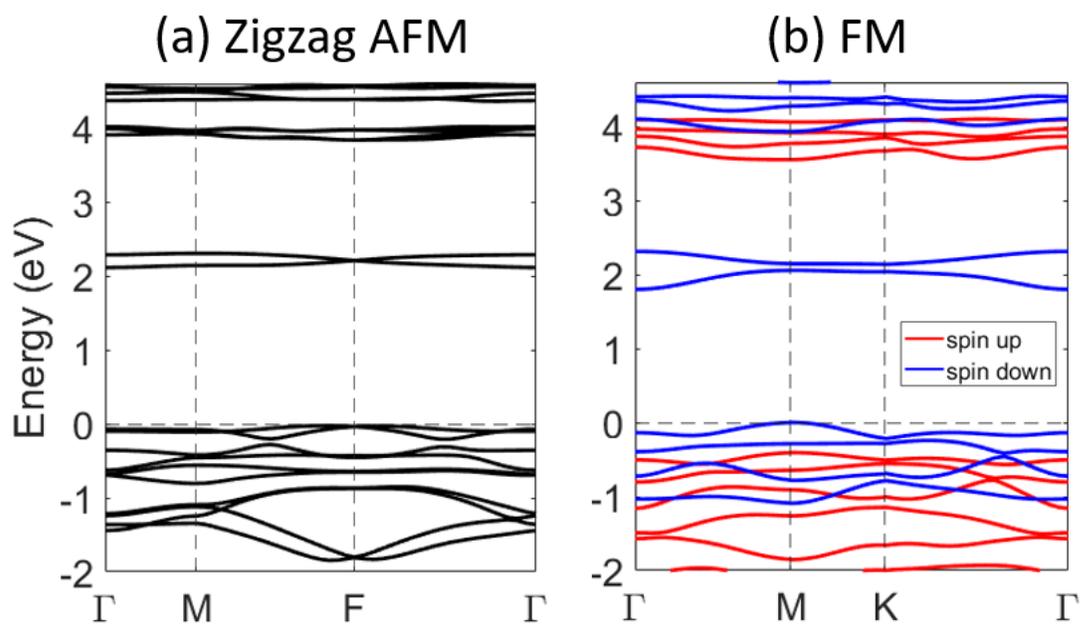

**FIG. 4.** Monolayer α-RuCl₃ QP electronic structure with (a) zigzag AFM and (b) FM phases.

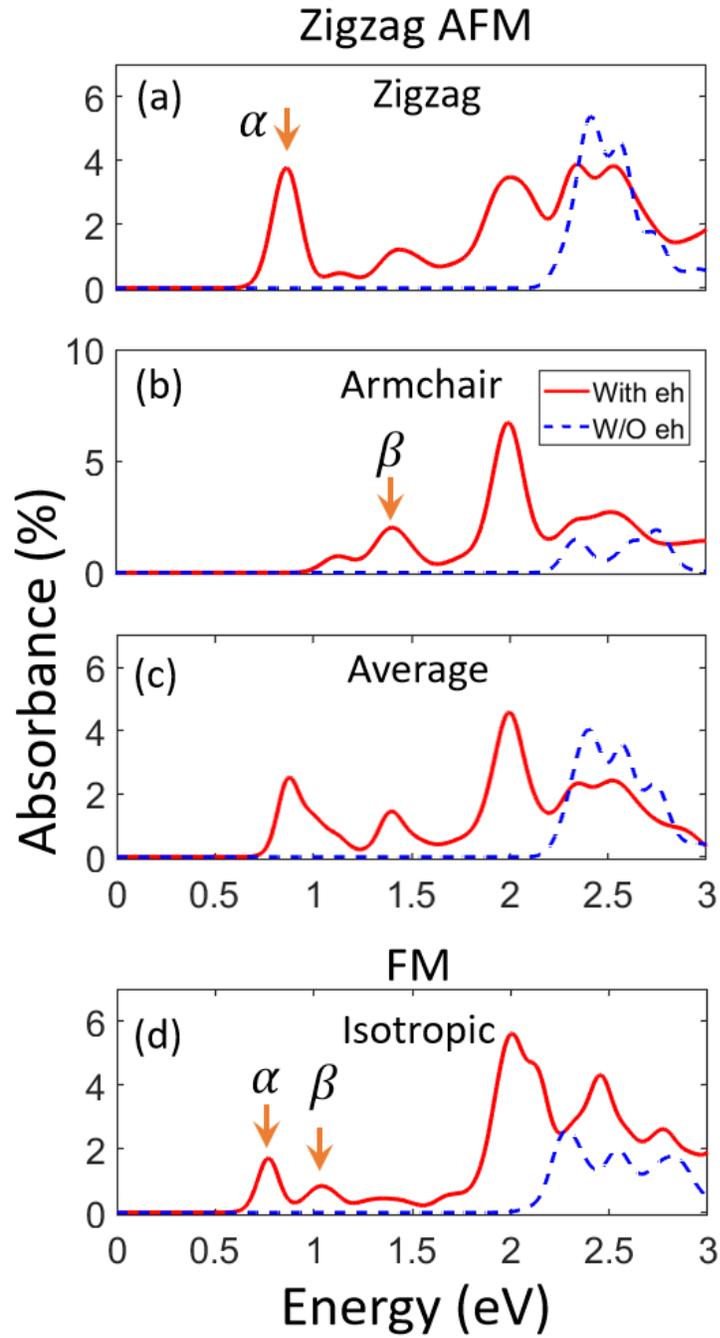

**FIG. 5.** Optical absorption spectrum of monolayer α-RuCl$_3$ without electron-hole interaction (blue dashed line) and with electron-hole interaction (red solid line) for (a-c) zigzag AFM and (d) FM phases. The incident light is polarized in (a) zigzag direction (b) armchair direction. (c) averaging all directions.

**TABLE**:

|  |  | Band Gap (eV) | | | Exciton Energy (eV) | | | | |
|---|---|---|---|---|---|---|---|---|---|
|  |  | DFT | QP | Expt. | $X_0$ | $\alpha$ | $\alpha$ expt. | $\beta$ | $\beta$ expt. |
| Bulk | Zigzag AFM | 0.78 (0.77) | 1.77 (1.75) | 1.80 ± 0.20 [19,20] | 1.14 | 1.23 | 1.16 [23,24] | 1.98 | 2.00 [23,24] |
| ML | Zigzag AFM | 0.90 (0.88) | 2.21 (2.20) | ~ 2 [13] | 0.76 | 0.86 | - | 1.41 | - |
| | FM | 0.80 (0.56) | 2.06 (1.84) | | 0.69 | 0.77 | - | 1.04 | - |

**TABLE I.** Summary of DFT and QP band gaps (the values listed are for the direct band gap, and the values in the parenthesis are for the indirect band gap), and exciton energies for bulk and monolayer $RuCl_3$.

**Notes**

The authors declare no competing financial interest.

**Data Availability Statement**

The data that support the findings of this study are available from the authors upon reasonable request.